\documentclass[twocolumn,prr,superscriptaddress,longbibliography]{revtex4-2}
\usepackage{physics}
\usepackage{graphicx}
\usepackage{here}
\usepackage{amsmath}
\usepackage{amssymb}
\usepackage{bm}
\usepackage{ascmac}
\usepackage{mathrsfs}
\usepackage{color}
\usepackage{comment}
\newcommand{\nn}{\nonumber \\}
\newcommand{\dg}{\dagger}
\newcommand{\im}{{\rm i}}
\newcommand{\h}{\hat}

\newcommand{\la}{\langle}
\newcommand{\ra}{\rangle}

\usepackage{hyperref}
\usepackage[normalem]{ulem}
\hypersetup{colorlinks=true,citecolor=blue,linkcolor=blue,urlcolor=blue}

\begin{document} 
\title{Evolution of entanglement entropy in strongly correlated bosons in an optical lattice} 
\author{Shion Yamashika}
\email{shion8ma4ka@icloud.com}
\affiliation{Department of Physics, Chuo University, Bunkyo, Tokyo 112-8551, Japan} 
\author{Daichi Kagamihara}
\email{kagamihara@phys.kindai.ac.jp}
\affiliation{Department of Physics, Kindai University, Higashi-Osaka, Osaka 577-8502, Japan}
\author{Ryosuke Yoshii}
\email{ryoshii@rs.socu.ac.jp}
\affiliation{Center of Liberal Arts and Sciences, Sanyo-Onoda City University, Yamaguchi 756-0884, Japan}
\author{Shunji Tsuchiya}
\email{tsuchiya@phys.chuo-u.ac.jp}
\affiliation{Department of Physics, Chuo University, Bunkyo, Tokyo 112-8551, Japan}
\date{\today}

\begin{abstract} 
We investigate the time evolution of the second-order Rényi entropy (RE) for bosons in a one-dimensional optical lattice following a sudden quench of the hopping amplitude $J$. Specifically, we examine systems that are quenched into the strongly correlated Mott-insulating (MI) regime with $J/U\ll 1$ ($U$ denotes the strength of the on-site repulsive interaction) from the MI limit with $J=0$.
In this regime, the low-energy excited states can be effectively described by fermionic quasiparticles known as doublons and holons. They are excited in entangled pairs through the quench dynamics. 
By developing an effective theory, we derive a direct relation between the RE and correlation functions associated with doublons and holons. This relation allows us to analytically calculate the RE and obtain a physical picture for the RE, both in the ground state and during time evolution through the quench dynamics, in terms of doublon-holon pairs.
In particular, we show that the RE is proportional to the population of doublon-holon pairs that span the boundary of the subsystem.
Our quasiparticle picture introduces some remarkable features that are absent in previous studies on the dynamics of entanglement entropy in free-fermion models.
It provides with valuable insights into the dynamics of entanglement entropy in strongly-correlated systems.
\end{abstract}

\maketitle

\section{Introduction}

Entanglement is one of the most intriguing concepts of quantum mechanics. It describes non-local correlations incompatible with local realism \cite{Einstein-1935}, which is clearly demonstrated by the violation of the Bell inequality \cite{Bell-1964}. 
Entanglement is also a key to understanding quantum many-body systems in diverse fields. It is considered, for example, to be the origin of thermalization in an isolated quantum many-body system \cite{Deutsch-1991,Srednicki-1994,Tasaki-1998,Kinoshita-2006,Rigol-2008,Yoshii-2022} and the Hawking radiation from black holes \cite{Hawking-1974,Hawking-1975}. 
In particular, entanglement plays a central role in strongly correlated systems. The investigation of entanglement in strongly correlated systems is promised to give us deep insights into fundamental aspects of emergent phenomena, such as quantum phase transition and topological order\,\cite{Kitaev2006,Levin2006,LAFLORENCIE20161,Osborne2002,Amico2008}.
\par
Entanglement between quantum objects can be quantified by entanglement entropy. 
It has been a major subject of theoretical investigation in quantum field theory, as well as in strongly correlated systems. Dynamics of entanglement entropy in integrable systems have been intensively investigated since the pioneering work by Calabrese and Cardy \cite{Calabrese-2005}. 
They proposed a clear physical picture for the dynamics of entanglement entropy in terms of quasiparticles. Specifically, the long-time dynamics of entanglement entropy can be understood as a result of excitation and propagation of entangled quasiparticle pairs. This quasiparticle picture has been confirmed numerically and analytically in a number of papers \cite{Fagotti-2008,Eisler-2008,Nezhadhaghighi-2014,Coser-2014,Cotler-2016,Buyskikh-2016,De-2006,Lauchli-2008,Kim-2013,Alba-2017,Alba-2017-2,Bertini-2022}.
\par
Despite recent developments of experimental techniques, measuring entanglement entropy remains challenging in condensed matter systems. A great advance has been recently made, however, in the system of ultracold bosonic atoms in an optical lattice. The second-order R\'{e}nyi entropy (RE), which is one of the measures of entanglement entropy, has been successfully probed by preparing two independent copies of the same state, letting them interfere, and counting the number parity of atoms in one of the copies by an atomic gas microscope \cite{Zoller-2012,Islam-2015}. The time evolution of the RE after a sudden quench of atomic hopping has been observed in the superfluid (SF) regime by using this technique \cite{Kaufman-2016}. In the strongly correlated Mott insulating (MI) regime, on the other hand, spreading of correlations after a quantum quench has been studied theoretically and experimentally \cite{Cheneau-2012,Takasu-2020,Kaneko-2022}. However, quench dynamics of entanglement entropy has not been well investigated in this regime.
\par
In this paper, motivated by the current status of theory and experiment, we study the quench dynamics of entanglement entropy of bosons in a one-dimensional optical lattice. 
Our main focus is on the quench dynamics of the RE in the strongly correlated Mott insulating regime. The low-energy dynamics in this regime can be effectively described by fermionic quasiparticles known as a doublon and a holon, which correspond to an excess particle and a hole on top of the unit filling, respectively \cite{Cheneau-2012}. 
We develop an effective theory to derive an analytical expression for the time evolution of the RE after a quench of atomic hopping. Furthermore, we derive a direct relation between the RE and correlation functions for doublons and holons, which enables us to obtain a physical picture for the dynamics of the RE in terms of doublon-holon pairs. We find that the obtained quasiparticle picture is consistent with the one proposed by Calabrese and Cardy in the space-time scaling limit. Moreover, it exhibits remarkable features in both the short and long time scales that are absent in their picture.
\par
The organization of the paper is as follows: In Sec.~\ref{sec:BHM}, we explain the model and setup for the quench dynamics of bosons in an optical lattice and introduce the RE. In Sec.~\ref{sec:eff_theory}, we introduce the effective theory in the strongly correlated MI regime. 
In Sec.~\ref{sec:RE_vac}, we introduce the formalism to calculate the RE and study the RE for the ground state. In Sec.~\ref{sec:RE}, we study the time evolution of the RE after a quench. In Sec.~\ref{sec:QP}, we discuss the physical picture for the time evolution of the RE. In Sec.~\ref{sec:nth-order_RE}, we extend the analysis to study the $n$-th order RE. In Sec.~\ref{sec:validity_of_the_effective_theory}, we examine the validity of the effective theory. We finally summarize the paper in Sec.~\ref{sec:summary}. We set $\hbar = 1$ and the lattice constant unity throughout this paper. 

\section{Model and setup}\label{sec:BHM}

We consider bosons in a one-dimensional (1D) optical lattice at zero temperature. 
When the lattice potential is deep enough, the system is well described by the Bose-Hubbard model (BHM) \cite{Fisher-1989,Jaksch-1998,Sachdev-2011}
\begin{align}
\h H = -J \sum_j (\h b_j^\dg \h b_{j+1}+{\rm H.c.}) 
+ \frac{U}{2} \sum_j \h n_j (\h n_j-1),\label{eq:BHM}
\end{align}
where $\h b_j$ ($\h b_j^\dg$) denotes the annihilation (creation) operator of a boson on the $j$th site and $\h n_j = \h b_j^\dg \h b_j$ the number operator on the $j$th site. $J$ denotes the hopping amplitude between nearest-neighbor sites and $U>0$ the strength of the on-site repulsive interaction. We assume the periodic boundary condition.
\par
The BHM (\ref{eq:BHM}) exhibits a quantum phase transition between the SF and MI phases \cite{Batrouni-1990,Kashurnikov-1996,Kuhner-1998,Danshita-2011,Ejima-2011,Carrasquilla-2013,Konstantin-2016}: When the total number of bosons $N$ is commensurate with the number of total sites $L$, the ground state is a SF state for small $U/J$, while it is a MI state for large $U/J$. The SF-MI phase transition of the Kosterlitz-Thouless type occurs at $U/J\simeq3.28$ for unit filling ($N/L=1$) \cite{Danshita-2011,Ejima-2011,Carrasquilla-2013,Konstantin-2016}. The ground state is a SF state when $N$ is incommensurate with $L$ regardless of the value of $U/J$. 
\par
We suppose that the whole system consists of subsystems A and B. The RE for subsystem A is defined as \cite{Horodecki-2009}
\begin{align}
S_{\rm A} = - \ln[\tr_{\rm A}(\h \rho_{\rm A}^2)],\label{eq:RE_def}
\end{align}
where $\hat{\rho}_{\rm A} = \tr_{\rm B}(\hat{\rho})$ is the reduced density matrix for subsystem A and $\hat{\rho}$ is the density matrix for the whole system. $\tr_{\rm A(B)}$ stands for trace over subsystem A (B).
$\tr_{\rm A}(\h \rho_{\rm A}^2)$ quantifies the purity of the state $\h \rho_{\rm A}$ \cite{Nakahara-2008}: $\tr_{\rm A}(\h \rho_{\rm A}^2)=1$ if $\h \rho_{\rm A}$ is a pure state, while $\tr_{\rm A}(\h \rho_{\rm A}^2)<1$ if it is a mixed state.
When subsystems A and B have no entanglement, $\hat{\rho}_{\rm A}$ describes a pure state and we obtain $S_{\rm A}=0$.
When subsystems A and B are entangled, $\hat{\rho}_{\rm A}$ describes a mixed state and we obtain $S_{\rm A}>0$.
\par
We follow the quench protocol of the experiments~\cite{Cheneau-2012,Islam-2015,Kaufman-2016}.
Namely, atoms are initially localized one in each of the lattice sites.
At the initial time ($t=0$), tunneling of atoms is abruptly switched on by lowering the lattice depth and the state of the whole system $|\psi(t)\rangle$ evolves following the Hamiltonian (\ref{eq:BHM}) as $|\psi(t)\rangle = e^{-\im\h H t}|\psi_0\rangle$, where $|\psi_0\rangle$ is the initial state at $t=0$.
\par 
The initial state can be written as 
\begin{align}
\ket{\psi_0} = \prod_{j=1}^L \h b_j^\dg \ket{0}_j, \label{eq:psi_0}
\end{align}
where $\ket{\nu}_j$ ($\nu=0,1,2,\dots$) denotes the Fock state with $\nu$ atoms on the $j$th site. It corresponds to the ground state of the MI limit ($J/U=0$).
Since the initial state (\ref{eq:psi_0}) is a product state, $S_{\rm A}=0$ at $t=0$. $S_{\rm A}$ grows in time after the quench as tunneling of bosons creates entanglement between the subsystems. 

\section{Effective theory in the strongly-correlated Mott Insulating regime}\label{sec:eff_theory}

We assume that the lattice potential is slightly lowered and the value of $J/U$ is set in the strongly correlated MI regime ($J/U\ll 1$) at $t>0$.
The low-energy excited states in this regime can be described in terms of doublons and holons~\cite{Cheneau-2012,Barmettler-2012}. 
Such a weak perturbation associated with the quench involves only the low-energy excited states. 
As a result, the time evolution of the system after the quench is considered to be well described by the effective theory based on the doublon-holon description. 
\par
Introducing the fermionic doublon and holon annihilation (creation) operators,  $\hat d_j$ ($\hat d_j^\dagger$) and $\hat h_j$ ($\hat h_j^\dagger$), respectively, the Hamiltonian \eqref{eq:BHM} is approximately mapped to \cite{Cheneau-2012,Barmettler-2012}
\begin{align}
\hat{H} = \hat{P} \hat{H}_{\rm eff} \hat{P},
\label{eq:H_PHeffP}
\end{align}
where $\hat{H}_{\rm eff}$ is the Hamiltonian of the effective theory given by
\begin{align}
\h H_{\rm eff} 
=& 
-J \sum_j [ 2 \h d_j^\dg \h d_{j+1} + \h h_{j+1}^\dg \h h_j \nn
&+ \sqrt 2 (\h d_j^\dg \h h_{j+1}^\dg - \h h_j \h d_{j+1})+{\rm H.c.}] \nn
&+ \frac{U}{2} \sum_j (\h d_j^\dg \h d_j + \h h_j^\dg \h h_j),\label{eq:H_eff}
\end{align}
and $\hat{P} = \prod_{j} (1 - \hat{d}^{\dag}_{j} \hat{d}_{j} \hat{h}_{j}^{\dag} \hat{h}_{j})$ is the projection operator, which eliminates double occupancy of a doublon and a holon on the same site.
The derivation of Eqs.~(\ref{eq:H_PHeffP}) and (\ref{eq:H_eff}) is given in Appendix~\ref{App:Heff}.
\par
The projection operator $\hat{P}$ can be safely neglected in weakly excited states of the strongly correlated MI regime since the system can be considered as a dilute gas of doublons and holons and the possibility of their occupation on the same site is quite low.
\par
By the Fourier transform, $\h H_{\rm eff}$ can be written as  
\begin{align}
\h H_{\rm eff} 
=&
\sum_k [f_{d,k} \h { d}_k^\dg \h { d}_k - f_{h,k} \h { h}_{-k} \h { h}_{-k}^\dg
\nn
&- \im g_k(\h { d}_k^\dg \h {h}_{-k}^\dg - \h { h}_{-k} \h { d}_k)], \label{eq:H_eff_k_}
\end{align}
where
$f_{d,k}=U/2-4J \cos (k)$, $f_{h,k}=U/2 - 2J \cos (k)$, and $g_k= 2\sqrt 2 J\sin (k)$. Doublon and holon have energy gap $f_{d,k=0}=U/2-4J$ and $f_{h,k=0}=U/2-2J$, respectively.
Note that the initial state $\ket{\psi_0}$ in Eq.~\eqref{eq:psi_0} corresponds to the vacuum state of $\hat{d}_j$ and $\hat{h}_j$.
\par
The quadratic Hamiltonian $\hat H_{\rm eff}$ can be diagonalized by the Bogoliubov transformation 
\begin{gather}
\mqty( \h \gamma_{d,k} \\ \h \gamma_{h,-k}^\dg )
= 
\mqty( u_k & -\im v_k \\ -\im v_k & u_k ) 
\mqty( \h { d}_k \\ \h { h}_{-k}^\dg ), \label{eq:bogolon}
\end{gather}
where $\hat \gamma_{d,k}$ and $\hat\gamma_{h,k}$ denote the annihilation operators of quasiparticles, which we refer to as ``bogolons" hereafter.
$u_k$ and $v_k$ are given by
\begin{align}
u_k 
&=
\sqrt{\frac{1}{2}\qty(1+\frac{f_{d,k}+f_{h,k}}{\sqrt{(f_{d,k}+f_{h,k})^2+4g_k^2}})}
\nn
&=
1  + O[(J/U)^2],
\label{eq:u_k}
\\
v_k 
&= {\rm sgn}(k)
\sqrt{\frac{1}{2}\qty(1-\frac{f_{d,k}+f_{h,k}}{\sqrt{(f_{d,k}+f_{h,k})^2+4g_k^2}})}
\nn
&=
2\sqrt{2} (J/U) \sin(k) + O[(J/U)^2],
\label{eq:v_k}
\end{align}
where we expand $u_k$ and $v_k$ in terms of $J/U$ for later use.
Substituting Eq.~(\ref{eq:bogolon}) into Eq.~(\ref{eq:H_eff_k_}), $\h H_{\rm eff}$ is diagonalized as
\begin{gather}
\h H_{\rm eff} 
= \sum_k (\epsilon_{d,k} \h \gamma_{d,k}^\dg \h \gamma_{d,k} + \epsilon_{h,k} \h \gamma_{h,-k}^\dg \h \gamma_{h,-k}),  \label{eq:H_diag}
\end{gather}%
where the dispersions of bogolons are given by 
\begin{align}
\epsilon_{d,k} &=  -J \cos (k) + \frac{1}{2}\sqrt{[U-6J\cos (k)]^2+32J^2\sin^2 (k)},\label{eq:e_d}\\
\epsilon_{h,k} &=  J \cos (k) + \frac{1}{2}\sqrt{[U-6J\cos (k)]^2+32J^2\sin^2 (k)}.\label{eq:e_h}
\end{align}
They have energy gap $\epsilon_{d,k=0}=U/2-4J$ and $\epsilon_{h,k=0}=U/2-2J$.
The ground state $|{\rm vac}\ra$ that satisfies $\h \gamma_{d,k}|{\rm vac}\ra = \h \gamma_{h,k}|{\rm vac}\ra=0$ can be written as \cite{Schrieffer-2018}
\begin{align}
|{\rm vac} \ra 
= 
\prod_k [u_k+\im v_k \h { d}_k^{\dg} \h { h}_{-k}^\dg]|\psi_0\ra.\label{eq:vac}
\end{align}%
It implies that doublon-holon pairs are condensed in the ground state from its similarity with the BCS wave function \cite{Bardeen-1957}.
\par
The time-evolving state after the quench is given as
\begin{align}
|\psi(t)\ra =& e^{-\im \h H_{\rm eff}t}|\psi_0\ra \nn
=&
\prod_k 
[ u_k 
-\im v_k e^{-\im (\epsilon_{d,k}+\epsilon_{h,k})t} \h \gamma_{d,k}^\dg \h \gamma_{h,-k}^\dg ] \ket{\rm vac},
\label{eq:psi_t}
\end{align}
where we used $\ket{\psi_0} = \prod_k [u_k - \im v_k \hat{\gamma}_{d,k}^{\dg} \hat{\gamma}_{h,-k}^\dg ] \ket{\rm vac}$, which can be obtained from $\hat{d}_{k} \ket{\psi_0} = \hat{h}_{k} \ket{\psi_0} = 0$.
It shows that pairs of bogolons are excited by the quench. 
Equation~(\ref{eq:psi_t}) will be used to calculate the time evolution of the RE.
\par
In terms of $\hat{d}_k$ and $\hat{h}_k$, Eq.~(\ref{eq:psi_t}) can be written as 
\begin{align}
    |\psi(t)\rangle 
    =&
    \prod_k 
    \left[
    u_k^2+v_k^2e^{-\im(\epsilon_{d,k}+\epsilon_{h,k})t} 
    \right. 
    \nn
    &
    +
    \left.\im u_k v_k 
    (1-e^{-\im(\epsilon_{d,k}+\epsilon_{h,k})t})
    \hat{d}_k^\dag \hat{h}_{-k}^\dag
    \right]|\psi_0\rangle.
    \label{eq:psi_t_dh}
\end{align}
Equation~(\ref{eq:psi_t_dh}) indicates that doublons and holons are excited in pairs. The numbers of doublons and holons should be thus equal: $\sum_{j=1}^L\bra{\psi(t)}\h d_j^\dagger \h d_j\ket{\psi(t)}=\sum_{j=1}^L\bra{\psi(t)}\h h_j^\dagger \h h_j\ket{\psi(t)}$.
Using this relation, the total number of original bosons is conserved as
\begin{align}
    \sum_{i=1}^L \langle \psi(t)|\hat n_i|\psi(t)\rangle 
    &= N+
    \sum_{i=1}^L 
    \langle \psi(t)|
    (\hat{d}_i^\dag \hat{d}_i-
    \hat{h}_i^\dag \hat{h}_i)
    |\psi(t)\rangle \nonumber \\
    &=N, 
\end{align}
where $N$ is the total number of original bosons at $t=0$. Note that the number of doublons (holons) itself is not conserved because they are excited from the vacuum state $|\psi_0\rangle$.

\section{R\'enyi entropy for the ground state}\label{sec:RE_vac}

Let us first calculate the RE for the ground state Eq.~(\ref{eq:vac}) before studying its time evolution.
For a Gaussian state, which includes the ground state and a thermal state of a quadratic Hamiltonian, the RE can be conveniently evaluated using single-particle correlation functions~\cite{Frerot-2015}. We first adopt the formalism to our system in Sec.~\ref{sec:RE_gaussian}, and apply it to the ground state in Sec.~\ref{sec:RE_deep_MI}.

\subsection{R\'enyi entropy for a Gaussian state}\label{sec:RE_gaussian}

We consider a Gaussian state of doublons and holons $\ket{\phi}$ with which any correlation function of $\h d_j$ and $\h h_j$ factorizes according to the prescriptions of Wick's theorem.
The reduced density matrix of $|\phi\rangle$ can be formally written as \cite{Frerot-2015}
\begin{align}
\h \rho_{\rm A} = \tr_{\rm B}(|{\phi}\ra\la {\phi}|)=
\frac{e^{-\h {\cal H}_{\rm A}}}{\tr_{\mathrm{A}} (e^{-\h {\cal H}_{\rm A}})}.
\label{eq:RDM_vac_main}
\end{align}
The entanglement Hamiltonian $\h {\cal H}_{\rm A}$ has a quadratic form of $\h d_j$ and $\h h_j$ ($j\in {\rm A}$), because Wick's theorem also holds for correlation functions concerning degree of freedom in {\rm A}. Thus, it can be diagonalized as
\begin{align}
\h {\cal H}_{\rm A}= \sum_{\alpha=1}^{2L_{\rm A}} \omega_{\alpha}^{\rm A} \h n_{\alpha}^{\rm A}, \label{eq:ent_Ham_vac}
\end{align}
where $\omega_{\alpha}^\mathrm{A}$ and $\h n_{\alpha}^\mathrm{A}$ are the spectrum and the number operator for the eigenmode $\alpha$. $L_\mathrm{A}$ is the size of subsystem A. 
Note that the number of the eigenmodes $2L_{\rm A}$ corresponds to the total number of degrees of freedom in subsystem A. 
Once the entanglement Hamiltonian is diagonalized in the form of Eqs.~\eqref{eq:ent_Ham_vac}, the RE can be obtained as
\begin{align}
S_{\rm A}= 
-\sum_{\alpha=1}^{2L_{\rm A}} \ln [(1-f_{\alpha})^2 + f_{\alpha}^2],
\label{eq:S_A_vac_main}
\end{align}
where $f_{\alpha}=\tr_{\rm A}(\h \rho_{\rm A} \h n_{\alpha}^\mathrm{A}) = 1/[\exp(\omega_{\alpha}^{\rm A})+1]$ is the occupation number of the eigenmode $\alpha$.  
\par
We still need to determine $\omega_{\alpha}$ or $f_{\alpha}$.
To obtain $f_\alpha$, we consider a matrix of single-particle correlation functions
\begin{align}
M=
\begin{pmatrix}
I - C_{d,d}^*& -C_{d,h}^* & F_{d,d} & F_{d,h} \\
-C_{h,d}^* & I- C_{h,h}^* & F_{h,d} & F_{h,h} \\
-F_{d,d}^* & -F_{d,h}^* & C_{d,d} & C_{d,h} \\
-F_{h,d}^* & -F_{h,h}^* & C_{h,d} & C_{h,h}
\end{pmatrix}.
\label{eq:M_vac_main}
\end{align}
Here, $C_{\sigma,\sigma'}$ and $F_{\sigma,\sigma'}$ ($\sigma,\sigma'=d,h$) are $L_{\rm A} \times L_{\rm A}$ matrices of normal and anomalous correlation functions, respectively, that have matrix elements
\begin{align}
(C_{\sigma,\sigma'})_{i,j} = \la\phi| \h c_{i,\sigma}^\dg \h c_{j,\sigma'}|\phi \ra,\label{eq:C_element_main}\\
(F_{\sigma,\sigma'})_{i,j} = \la \phi|\h c_{i,\sigma} \h c_{j,\sigma'}|\phi\ra,\label{eq:F_element_main}
\end{align}
where we denote $\h c_{i,d} = \h d_i$ and $\h c_{i,h}=\h h_i$. 
$f_{\alpha}$ can be obtained by diagonalizing $M$ thanks to the relation \cite{Frerot-2015}
\begin{align}
M= 
U_{\rm A}
\mqty( {\rm diag} (1-f_{\alpha}) & 0 \\ 0 & {\rm diag}(f_{\alpha}) )
U_{\rm A}^{-1}, \label{eq:M_diag_main}
\end{align}
where $U_{\rm A}$ is a unitary matrix.

\subsection{R\'enyi entropy for the ground state}
\label{sec:RE_deep_MI}

We calculate the RE for the ground state Eq.~(\ref{eq:vac}) employing the formalism in Sec.~\ref{sec:RE_gaussian}.
Evaluating the single-particle correlation functions in Eqs.~(\ref{eq:C_element_main}) and (\ref{eq:F_element_main}) with $|{\rm vac}\rangle$, we obtain $C_{d,h}=C_{h,d}=0$, $F_{d,d}=F_{h,h}=0$, $(C_{d,d})_{i,j}=(C_{h,h})_{i,j}= C_{i,j}$, and $(F_{d,h})_{i,j}=(F_{h,d})_{i,j}= F_{i,j}$, where
\begin{align}
C_{i,j}=\bra{{\rm vac}}\hat d_i^\dagger\hat d_j\ket{{\rm vac}}=\bra{{\rm vac}}\hat h_i^\dagger\hat h_j\ket{{\rm vac}}\notag
\\=
\frac{1}{L}\sum_k
v_k^2 e^{\im k(i-j)},\label{eq:C_vac}
\\
F_{i,j}=\bra{{\rm vac}}\hat d_i\hat h_j\ket{{\rm vac}}=\bra{{\rm vac}}\hat h_i\hat d_j\ket{{\rm vac}}\notag
\\=
\frac{\im}{L}\sum_k
u_k v_k
e^{\im k(i-j)}. 
\label{eq:F_vac}
\end{align}
From $u_k=O(1)$ and $v_k=O(J/U)$ in Eqs.~(\ref{eq:u_k}) and (\ref{eq:v_k}), we obtain $C_{i,j}=O[(J/U)^2]$ and $F_{i,j}=O(J/U)$, because the summations over $k$ in Eqs.~(\ref{eq:C_vac}) and (\ref{eq:F_vac}) do not change the order of $J/U$.
\par
Using matrix $C$ and $F$, the matrix of the single-particle correlation functions $M$ in Eq.~(\ref{eq:M_vac_main}) can be simplified as
\begin{gather}
M
=
\mqty(
I-C &0 & 0& F \\
0 & I-C & F & 0 \\
0 & -F^* & C & 0 \\
-F^* & 0 & 0 & C
). \label{eq:M_vac_2}
\end{gather}
Here, $F$ is antisymmetric, i.e., $F_{i,j}=-F_{j,i}$, due to the anticommutation relation $\{\hat d_i,\hat h_j\}=0$. Note that this property holds for any state with which the average is taken.
\par
We derive a formula that directly relates the RE and the single-particle correlation functions.
From Eq.~\eqref{eq:M_vac_2}, we obtain 
\begin{align}
\tr(M^2)
&= 
2L_{\rm A} - 4\tr(C) +4\norm{F}_{\rm F}^2
+O[(J/U)^4],
\label{eq:M_vac_tr}
\end{align}
where $\norm{O}_{\rm F}=\sqrt{\sum_{i,j}|O_{i,j}|^2}$ denotes the Frobenius norm. 
In deriving Eq.~\eqref{eq:M_vac_tr}, we use $C=O[(J/U)^2]$ and $\tr(FF^*)=-\norm{F}_\mathrm{F}^2$, which is obtained from $F_{i,j}=-F_{j,i}$.
On the other hand, using Eq.~(\ref{eq:M_diag_main}), we obtain
\begin{align}
\tr(M^2)
&= 
2L_{\rm A}- 2\sum_{\alpha=1}^{2L_{\rm A}} (f_\alpha- f_\alpha^2).
\label{eq:M_vac_tr_2}
\end{align}
Comparing Eqs.~(\ref{eq:M_vac_tr}) and (\ref{eq:M_vac_tr_2}), we obtain a relation between $f_\alpha$ and the correlation functions as
\begin{align}
\sum_{\alpha=1}^{2L_{\rm A}} (f_\alpha - f_\alpha^2)=
2 \qty[\tr(C)- \norm{F}_{\rm F}^2]+O[(J/U)^4].
\label{eq:f_vac}
\end{align}
The first term of the right-hand side in Eq.~(\ref{eq:f_vac}) is of the order of $(J/U)^2$ because $C=O[(J/U)^2]$ and $F=O(J/U)$.
Since $f_{\alpha} - f_{\alpha}^2 \geq 0$, we obtain $f_{\alpha} - f_{\alpha}^2 = O[(J/U)^2]$ from Eq.~\eqref{eq:f_vac}.
Expanding Eq.~\eqref{eq:S_A_vac_main} by $f_{\alpha} - f_{\alpha}^2$ and using Eq.~\eqref{eq:f_vac}, the RE can be obtained in a concise form:
\begin{align}
S_{\rm A}
= 
4\qty[\tr(C)- \norm{F}_{\rm F}^2]+O[(J/U)^4].
\label{eq:RE_vac_}
\end{align}
Remarkably, Eq.~\eqref{eq:RE_vac_} allows us to gain a clear quasiparticle picture for the RE, which we will discuss in Sec.~\ref{sec:QP}.
\par
In deriving the above formula, we use only $C=O[(J/U)^2]$ and $F=O(J/U)$ in addition to the relation $F_{i,j}=-F_{j,i}$, which holds for any state.
As long as these conditions are satisfied, therefore, Eq.~(\ref{eq:RE_vac_}) holds for any Gaussian state, regardless of the explicit forms of $C$ and $F$.
We will take advantage of this fact to calculate the RE for the time-evolving state.
\par
In the limit $L\to \infty$, replacing the summations over $k$ with integrals [$\sum_k\to(L/2\pi)\int_{-\pi}^\pi dk$] in Eqs.~(\ref{eq:C_vac}) and (\ref{eq:F_vac}), we obtain
\begin{align}
    C_{i,i} =& 4(J/U)^2 +O[(J/U)^4], \label{eq:C_ii_vac}
    \\
    F_{i,j} =& \mp \sqrt{2}(J/U) \delta_{i,j\pm1} + O[(J/U)^2].
    \label{eq:F_ij_vac}
\end{align}
\par
Substituting Eqs.~(\ref{eq:C_ii_vac}) and (\ref{eq:F_ij_vac}) into Eq.~(\ref{eq:RE_vac_}), the RE is obtained as
\begin{align}
    S_{\rm A}
    = 16(J/U)^2 +O[(J/U)^3].
    \label{eq:RE_vac}
\end{align}
The above expression clearly shows that $S_{\rm A}$ is independent of subsystem size $L_{\rm A}$ and therefore follows the {\it area-law scaling}, which is a characteristic feature of a gapped ground state of a short-range Hamiltonian \cite{Eisert-2010}. 
The ground state $\ket{\rm vac}$ indeed satisfies this condition.
\par
\begin{figure}[t!]
\includegraphics[width=0.45\textwidth]{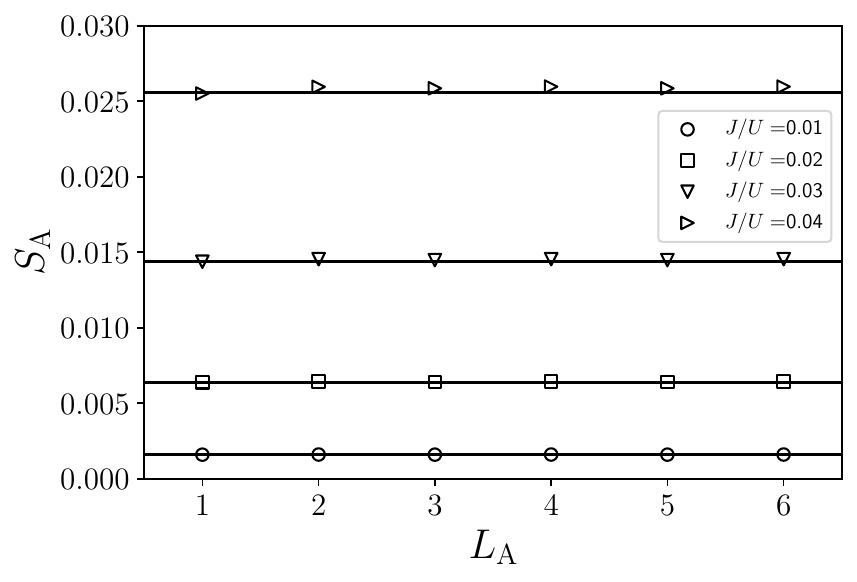}
\caption{RE for the ground state $|{\rm vac}\rangle$.
The analytic results in Eq.~(\ref{eq:RE_vac}) are indicated by the horizontal solid lines.
The symbols indicate the numerical data of the iTEBD calculation. }
\label{fig:RE_vac}
\end{figure}

Figure~\ref{fig:RE_vac} shows a comparison of Eq.~(\ref{eq:RE_vac}) with the numerical results obtained by the matrix-product-state technique. 
They agree well with each other.
To obtain the numerical results, we calculate the ground state wave function of the BHM (\ref{eq:BHM}) by imaginary-time evolution using the infinite time-evolving block-decimation (iTEBD) algorithm \cite{Orus-2008} and evaluate the RE with it.
In the numerical calculations using the iTEBD algorithm throughout this paper, we implement it keeping the Schmidt coefficients larger than $10^{-7}$, setting the dimension of the local Hilbert space being 5, and using the second-order Suzuki-Trotter decomposition with time step $\Delta t =0.01/U$.

\section{Time-evolution of the R\'enyi entropy}\label{sec:RE}

In this section, we study the time evolution of the RE. We calculate the RE for a single site ($L_{\rm A}=1$) in Sec.~\ref{sec:single_site} by directly evaluating the reduced density matrix.
We further calculate the RE for $L_{\rm A}\geq 1$ in Sec.~\ref{sec:RE_L_A>1} using the formalism in 
Sec.~\ref{sec:RE_vac}.

\subsection{R\'enyi entropy for a single site}\label{sec:single_site}

The reduced density matrix for subsystem A can be written in the basis of the Fock states of doublons and holons $|n_d,n_h\rangle_j$ ($n_d,n_h=0,1$) as
\begin{align}
\hat \rho_\mathrm{A}
= 
\sum_{\tau,\tau'\in\{(0,0),(1,0),(0,1),(1,1)\}}
r_{\tau,\tau'} |\tau \rangle_j \langle \tau'|_j,
\end{align}
where $n_d$ ($n_h$) denotes the number of doublon (holon) on the $j$th site.
The matrix elements are given as
\begin{align}
r_{\tau,\tau'} =& \tr_{\mathrm{A}} \left[ \hat{\rho}_{\mathrm{A}} |\tau' \rangle_j \langle \tau|_j \right]
= \tr \left[ \hat{\rho} |\tau' \rangle_j \langle \tau|_j \otimes \hat{1}_{\bar{j}} \right]
\nonumber
\\
=& \langle \psi(t) | \left[ | \tau' \rangle_j \langle \tau|_j \otimes \hat{1}_{\bar{j}} \right] |\psi(t) \rangle,
\label{eq:rdm_tomography}
\end{align}
where $\hat{1}_{\bar{j}}$ denotes the identity operator for all the sites other than $j$.
\par
The matrix elements can be calculated as
\begin{align}
\hat{\rho}_{\mathrm{A}}
=&
[1 - 2r(t) + r(t)^2] |0,0 \rangle_j \langle 0,0|_j
\nonumber \\
&+
\left[ r(t) - r(t)^2 \right] \left(  |1,0 \rangle_j \langle 1,0|_j + |0,1 \rangle_j \langle 0,1|_j \right)
\nonumber \\
&+
r(t)^2 |1,1 \rangle_j \langle 1,1|_j,
\label{eq:rdm_effective_explicit}
\end{align}
where $r(t) = \langle \psi(t) | \hat{d}^{\dag}_{j} \hat{d}_{j} |\psi(t) \rangle = \langle \psi(t) | \hat{h}^{\dag}_{j} \hat{h}_{j} |\psi(t) \rangle$ represents the number of doublon (holon) per site
\begin{align}
r(t)=\frac{2}{L} \sum_k u_k^2v_k^2 \qty{1-\cos \qty[(\epsilon_{d,k}+\epsilon_{h,k})t]}.
\label{eq:r_t}
\end{align}
The calculation of matrix elements $r_{\tau,\tau'}$ is  straightforward once we express the operators $|\tau \rangle_j \langle \tau'|_j$ in terms of the creation and annihilation operators of doublons and holons. 
The detail of the calculation is given in Appendix~\ref{app:rho_A_single}.
\par
Substituting Eq.~\eqref{eq:rdm_effective_explicit} into Eq.~\eqref{eq:RE_def} and noting that $r(t) = O[(J/U)^2]$, we find that the RE is proportional to the number of doublon (holon) per site in the leading order of $J/U$ as
\begin{align}
S_{\rm A} =& -\ln \left\{[1-2r(t) + r(t)^2]^2 +  2[r(t) - r(t)^2]^2 + r(t)^4 \right\}
\nonumber \\
=& 4r(t) + O[(J/U)^4]. \label{eq:RE_eff}
\end{align}

\begin{figure}[!t]
\includegraphics[width=0.5\textwidth]{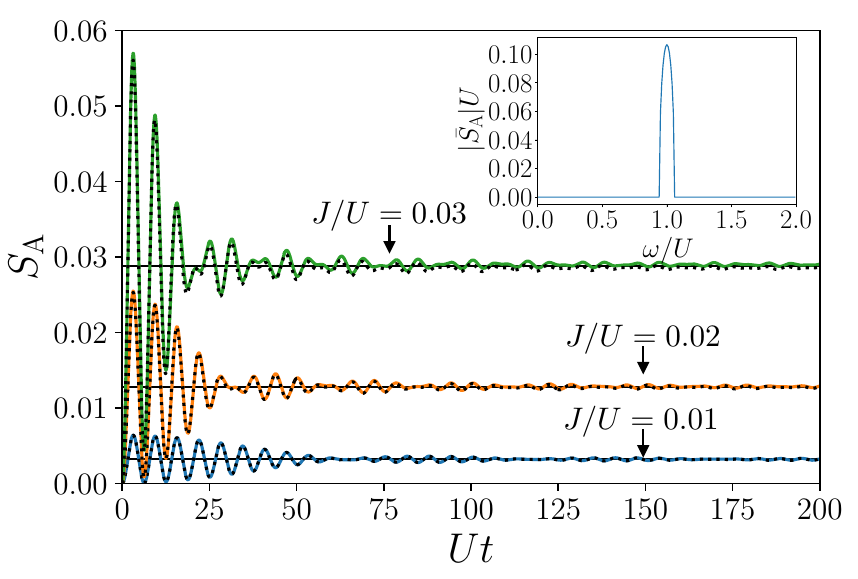}
\caption{Time evolution of the RE for a single site ($L_A=1$).
Equation~(\ref{eq:S_Bessel}) is plotted as the solid lines.
The horizontal lines indicate the asymptotic values $S_{\rm A}(t\to\infty)=32(J/U)^2$. The dotted lines indicate the numerical results of the iTEBD calculations. The inset shows the Fourier transform $\bar{S}_{\rm A}(\omega)$ in Eq.~(\ref{eq:S_Bessel_omg}) for $J/U =0.01$.}
\label{fig:RE_relax}
\end{figure}

To figure out why the RE is proportional to the number of doublon (holon) per site,
we expand the wave function (\ref{eq:psi_t}) to the first order of $J/U$ as
\begin{align}
|\psi(t)\ra =&
|\psi_0\ra 
+ 2\im (J/U) \sum_k \sin(k)
[1-e^{-\im(\epsilon_{d,k}+\epsilon_{h,k})t }]
\nn
& \times 
\frac{\h d_k^\dg \h h_{-k}^\dg+\h h_{k}^\dg\h d_{-k}^\dg }{\sqrt 2}  |\psi_0\ra
+ O[(J/U)^2]. \label{eq:psi_t_expand}
\end{align}
The second term illustrates that all the doublons and holons are excited in {\it entangled pairs} \cite{Cheneau-2012}. 
Recalling that the RE quantifies entanglement between subsystems A and B, entangled doublon-holon pairs spanning the boundary between subsystems A and B should contribute to the RE.
Since the number of doublons (holons) of subsystem A (site $j$) is equal to that of doublon-holon pairs spanning subsystems A and B, the RE is naturally proportional to the number of doublons (holons).
\par
In the limit $L\rightarrow \infty$, evaluating the summation in Eq.~(\ref{eq:r_t}) by replacing it with an integral, we obtain 
\begin{equation}
S_{\rm A}(t) = 32(J/U)^2
\qty[
1- \frac{{\cal J}_1(6Jt)}{3Jt} \cos (Ut)
] + O[(J/U)^4], \label{eq:S_Bessel} 
\end{equation}
where ${\cal J}_n(x)$ ($n=0,1,2,\cdots$) is the Bessel function of the first kind. The density of doublon-holon pairs is indeed $r(t)=S_{\rm A}(t)/4$.
\par
Figure~\ref{fig:RE_relax} shows the RE in Eq.~\eqref{eq:S_Bessel} as a function of time.
The RE rapidly oscillates soon after the quench and converges after a while.
The constant value after the convergence increases as $J/U$ increases.
Equation~(\ref{eq:S_Bessel}) indicates that the frequency of the oscillation is equivalent to $U$, and $S_{\rm A}$ converges in the time scale of $O(1/J)$ because $\mathcal{J}_1(6Jt)/(Jt) = O[(Jt)^{-3/2}]\ll 1$ for $Jt\gg 1$.
The constant after the convergence is given as
\begin{equation}
\lim_{t \rightarrow \infty} S_{\rm A}(t) = 32 (J/U)^2 + O[(J/U)^4]\label{eq:S_limit}.
\end{equation}%
Figure~\ref{fig:RE_relax} shows that Eqs.~(\ref{eq:S_Bessel}) and (\ref{eq:S_limit}) agree well with the RE calculated by the iTEBD algorithm.
\par
The Fourier transform of Eq.~(\ref{eq:S_Bessel}) is given as
\begin{align}
\bar S_{\rm A} (\omega)
&= \int_{-\infty}^{\infty} \dd t~ S_{\rm A}(t) e^{\im \omega t} 
\label{eq:S_omg_1}
\\
&= 
-\frac{32J}{3U^2} \sqrt{1-\qty(\frac{U-\omega}{6J})^2} 
\theta(6J-|\omega-U|) 
\nn
&\quad+O[(J^3/U^4)],
\label{eq:S_Bessel_omg}
\end{align}
where $\theta(x)$ is the step function.
We assume that the frequency $\omega$ is positive to eliminate the contribution from the time-independent term in Eq.~(\ref{eq:S_Bessel}).
We also assume for simplicity that $S_{\mathrm{A}}(t) = S_{\mathrm{A}}(-t)$ for $t<0$.
The detail of the derivation of Eq.~(\ref{eq:S_Bessel_omg}) is given in Appendix~\ref{app:Fourier}. $\bar S_\mathrm{A}(\omega)$ has a peak at $\omega=U$ and its width is $\Delta\omega=12J$, as shown in the inset of Fig.~\ref{fig:RE_relax}.
The rapid oscillations with the frequency $U$ are induced by excited bogolons in Eq.~(\ref{eq:psi_t}). 
Their excitation energy can be approximated as $\epsilon_{d,k}+\epsilon_{h,k}\simeq U-6J\cos (k)$ for $J/U\ll 1$. 
The peak position $\omega=U$ corresponds to the center of the energy band, while the peak width $12J$ corresponds to its bandwidth.

\subsection{R\'enyi entropy for $L_{\rm A}\geq 1$}\label{sec:RE_L_A>1}
The reduced density matrix for the time-evolving state  $|\psi(t)\rangle$ under the quadratic Hamiltonian $\hat{H}_{\rm eff}$ can be formally written in the form of a thermal state as
\begin{equation}
\h \rho_{\rm A} = {\rm tr}_{\rm B}(|{\psi(t)}\ra \la \psi(t)|)=
\frac{e^{-\h {\cal H}_{\rm A}}}{\tr_{\mathrm{A}}(e^{-\h {\cal H}_{\rm A}})}, \label{eq:rho_A_quad}
\end{equation}
where $\h {\cal H}_{\rm A}(t)$ can be written as a quadratic form of $\h d_j$ and $\h h_j$ ($j \in {\rm A}$), because the Bloch-De Dominicis theorem holds for the correlation functions evaluated by $\h \rho_\mathrm{A}$ (see Appendix~\ref{App:rho_A}). 
It means that the time-evolving state is also a Gaussian state, and hence the RE can be calculated using the formalism in Sec.~\ref{sec:RE_gaussian}.
\par 
Evaluating the single-particle correlation functions in Eqs.~(\ref{eq:C_element_main}) and (\ref{eq:F_element_main}) with $|\psi(t)\ra$, we obtain $C_{d,h}=C_{h,d}=0$, $F_{d,d}=F_{h,h}=0$, $(C_{d,d})_{i,j}=(C_{h,h})_{i,j}=C_{i,j}$, and $(F_{d,h})_{i,j}=(F_{h,d})_{i,j}=F_{i,j}$, where
\begin{align}
C_{i,j}
=&
\la \psi(t)|\h d_i^\dg \h d_j|\psi(t)\ra
=
\la \psi(t)|\h h_i^\dg \h h_j|\psi(t)\ra,
\nn
=&
\frac{2}{L}\sum_k
u_k^2 v_k^2 
\{1
-\cos[(\epsilon_{d,k}+\epsilon_{h,k})t]
\}
e^{\im k (i-j)},
\label{eq:C_t}
\\
F_{i,j}
=&
\la \psi(t)|\h d_i \h h_j|\psi(t)\ra
=
\la \psi(t)|\h h_i \h d_j|\psi(t)\ra,
\nn
=&
\frac{\im}{L}\sum_k
u_k v_k e^{\im k(i-j)}
\nn
&\times
\qty[
v_k^2(1-e^{\im (\epsilon_{d,k}+\epsilon_{h,k})t})
-
u_k^2(1-e^{-\im (\epsilon_{d,k}+\epsilon_{h,k})t})
].
\label{eq:F_t}
\end{align}
Note that $C_{i,i}=r(t)$ and $F_{i,i}=0$. 
We find $C_{i,j}=O[(J/U)^2]$ and $F_{i,j}=O(J/U)$ from Eqs.~(\ref{eq:C_t}) and (\ref{eq:F_t}). Thus, the RE for the time-evolving state $|\psi(t)\rangle$ can be written in the same form as Eq.~(\ref{eq:RE_vac_}).
\par
In the limit $L\to\infty$, evaluating the summation in Eq.~(\ref{eq:F_t}), we obtain 
\begin{align}
F_{i,i+n}=&\sqrt{2}(J/U)(\delta_{n,1}-\delta_{n,-1})
\nn
&+\sqrt{2}(J/U)\im^{n+1}e^{-\im Ut}n\frac{{\cal J}_n(6Jt)}{3Jt}
+O[(J/U)^2],\label{eq:F_n}\\
\norm{F}_{\rm F}^2=&4(J/U)^2(L_{\rm A}-1)
\nn
&+4(J/U)^2\left\{\sum_{n=0}^{L_{\rm A}}(L_{\rm A}-n)n^2\left(\frac{{\cal J}_n(6Jt)}{3Jt}\right)^2\right.\nn
&\left.
-2(L_{\rm A}-1)\frac{{\cal J}_1(6Jt)}{3Jt}\cos(Ut)\right\}+O[(J/U)^3].\label{eq:|F|_F}
\end{align}
$C_{i,i}=r(t)$ is given in Eq.~\eqref{eq:r_t}.
Using~Eq.(\ref{eq:RE_vac_}), we obtain the RE for $L_{\rm A}\geq 1$ as 
\begin{align}
S_{\rm A}(t)
=& 
16\qty(\frac{J}{U})^2 (L_{\rm A}+1) 
-32 \qty(\frac{J}{U})^2 \cos(Ut)\frac{{\cal J}_1(6Jt)}{3Jt} 
\nn
&-16\qty(\frac{J}{U})^2\sum_{n=0}^{L_{\rm A}} (L_{\rm A}-n)n^2 
\qty(\frac{{\cal J}_n(6Jt)}{3Jt})^2 
\nn
&+O[(J/U)^3].
\label{eq:RE_analytic}
\end{align}
Setting $L_{\rm A}=1$, the above equation indeed reduces to Eq.~(\ref{eq:S_Bessel}). 

\begin{figure}[!t]
\includegraphics[width=0.5\textwidth]{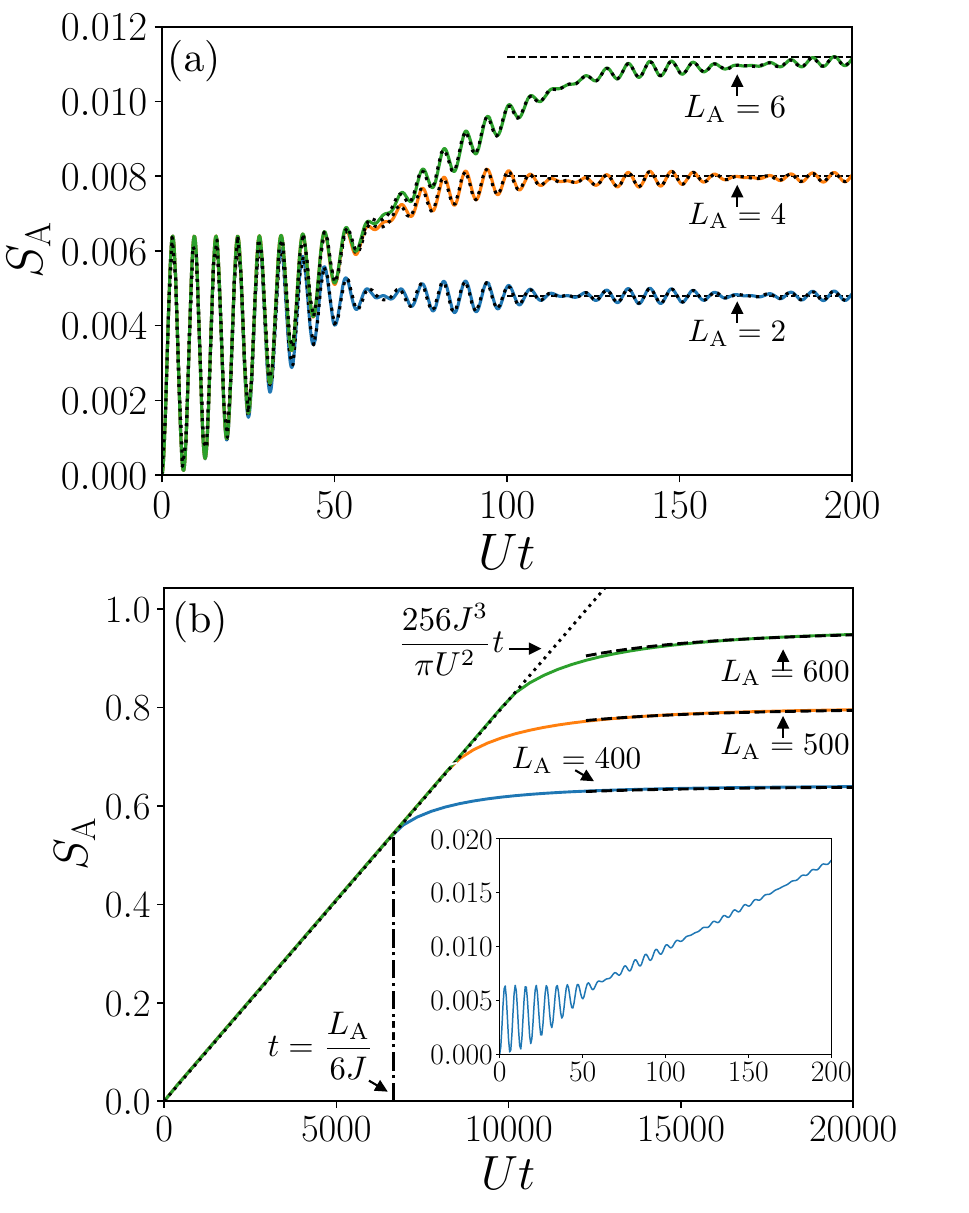}
\caption{Time evolution of the RE for $L_{\rm A}\geq 1$. We set $J/U=0.01$. Equation~(\ref{eq:RE_analytic}) is plotted as the solid lines.
(a) REs for $L_{\rm A}=2,4,$ and 6 in the short-time scale. The numerical results of the iTEBD algorithm are indicated by the dotted lines. The dashed lines indicate $16(J/U)^2(L_{\rm A}+1)$. (b) REs for $L_{\rm A}=400,500$, and 600 in the long-time scale. The dotted and dashed lines indicate the asymptotic forms of the RE in Eq.~(\ref{eq:RE_asymptotic}).
The inset shows a magnification in the short-time scale for $L_{\rm A}=400$. 
}
\label{fig:S_V_law}
\end{figure}

Figures~\ref{fig:S_V_law} (a) and (b) show $S_{\rm A}(t)$ in Eq.~(\ref{eq:RE_analytic}) as functions of time in the short-time scale for a small subsystem ($Jt,L_{\rm A}=O(1)$) and long-time scale for a large subsystem ($Jt, L_{\rm A}\gg 1$), respectively.
In the former, the numerical results of the iTEBD calculations agree well with Eq.~(\ref{eq:RE_analytic}), as shown in Fig.~\ref{fig:S_V_law} (a).
Analogous to Fig.~\ref{fig:RE_relax}, $S_{\rm A}(t)$ exhibits rapid oscillations with the frequency $U$ and converges in the time scale of $O(1/J)$.
Meanwhile, in Fig.~\ref{fig:S_V_law} (b), we find that $S_{\rm A}(t)$ linearly increases for $6Jt<L_{\rm A}$ and is saturated to a constant proportional to the size of the subsystem $L_{\rm A}$ for $6Jt>L_{\rm A}$. 
The RE after a long time thus obeys the {\it volume-law scaling}.
We can confirm these behaviors analytically in the asymptotic forms of Eq.~(\ref{eq:RE_analytic}) as
\begin{gather}
S_\mathrm{A}\simeq
\begin{cases}
     \displaystyle 
     \frac{256J^3}{\pi U^2}t, & 
     (1\ll Jt\ll L_\mathrm{A}), \\ \\
     \displaystyle
     \frac{16J^2(L_\mathrm{A}+1)}{U^2}-\frac{2L_\mathrm{A}^4}{81\pi J U^2 t^3},& (1\ll L_\mathrm{A}\ll Jt).
     \label{eq:RE_asymptotic}
\end{cases}
\end{gather}
The second asymptotic form indicates that the RE approaches to a constant with a correction of order $O(L_\mathrm{A}^4/t^3)$.
Note that the oscillations with frequency $U$ can be seen in the short-time scale even for a large subsystem, as shown in the inset of Fig.~\ref{fig:S_V_law} (b).

\section{Quasiparticle picture}\label{sec:QP}

\begin{figure}[!t]
\includegraphics[width=0.45\textwidth]{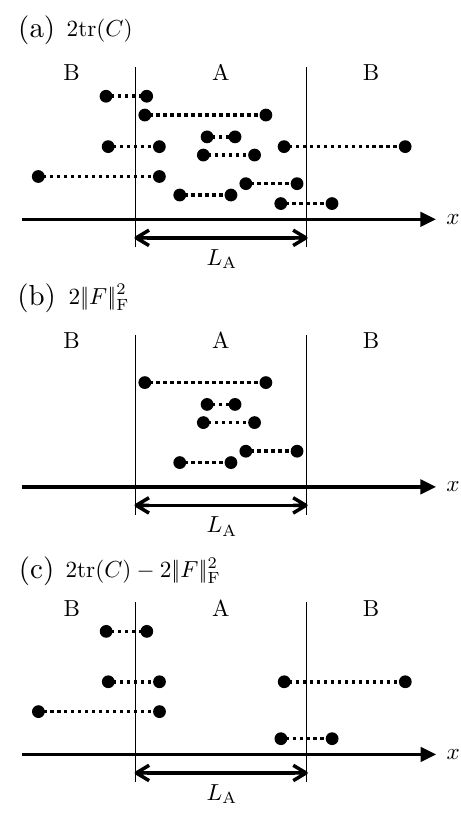}
\caption{Schematic drawings describe the physical image of (a) $\tr(C)$, (b) $\norm{F}^2$, and (c) $\tr(C)-\norm{F}^2$. Each two dots connected by a dotted line represents an entangled pair of a doublon and a holon.}
\label{fig:pair}
\end{figure}

In this section, we discuss how the RE in the ground state and the time-evolving state can be understood in terms of doublon-holon pairs.
First of all, recall that the RE for both $|\mathrm{vac}\ra$ and $|\psi(t)\ra$ is expressed as
\begin{align}
 S_\mathrm{A} \simeq 2\qty[2\tr(C)-2\norm{F}_\mathrm{F}^2],
 \label{eq:quasipic}
\end{align}
where we neglect $O[(J/U)^4]$.
The above expression can be understood as follows:
$2{\rm tr}(C)=\sum_{j \in {\rm A}} (\la \h n_{jd}\ra +\la \h n_{jh}\ra)$ represents the total number of doublons and holons in subsystem A, which is equal to the sum of the number of doublon-holon pairs spanning the boundary of subsystem A and twice the number of doublon-holon pairs within subsystem A [See Fig.~\ref{fig:pair} (a)]. 
Meanwhile, $\norm{F}_{\rm F}^2$ can be written as
\begin{equation}
\norm{F}_{\rm F}^2 
= \sum_{i,j \in \mathrm{A}} \langle \hat{h}^{\dag}_{j} \hat{d}^{\dag}_{i} \hat{d}_{i} \hat{h}_{j} \rangle.
\label{eq:pairnumber}
\end{equation}
Here, we neglect a correction of $O[(J/U)^4]$.
$2\norm{F}_{\rm F}^2$ is equal to twice the number of pairs within subsystem A [See Fig.~\ref{fig:pair} (b)].
$2{\rm tr}(C)-2\norm{F}_{\rm F}^2$ is thus equal to the number of doublon-holon pairs spanning the boundary of A [See Fig.~\ref{fig:pair} (c)]. Equation~(\ref{eq:quasipic}) indicates that they are responsible for entanglement between subsystems A and B, which is consistent with the fact that the doublon-holon pairs are entangled.

The area-law scaling of the ground state can be indeed understood within this quasiparticle picture. Given that $F_{i,j}$ is nonzero only when $|i-j|=\pm1$ in Eq.~(\ref{eq:F_ij_vac}), doublon-holon pairs spread between nearest-neighbor sites in the ground state. It turns out that only the pairs adjacent to the boundary can straddle the boundary and the number of such pairs does not depend on the size of the subsystem $L_\mathrm{A}$. Hence, the RE obeys the area-law scaling.

\begin{figure}[!t]
\includegraphics[width=0.52\textwidth]{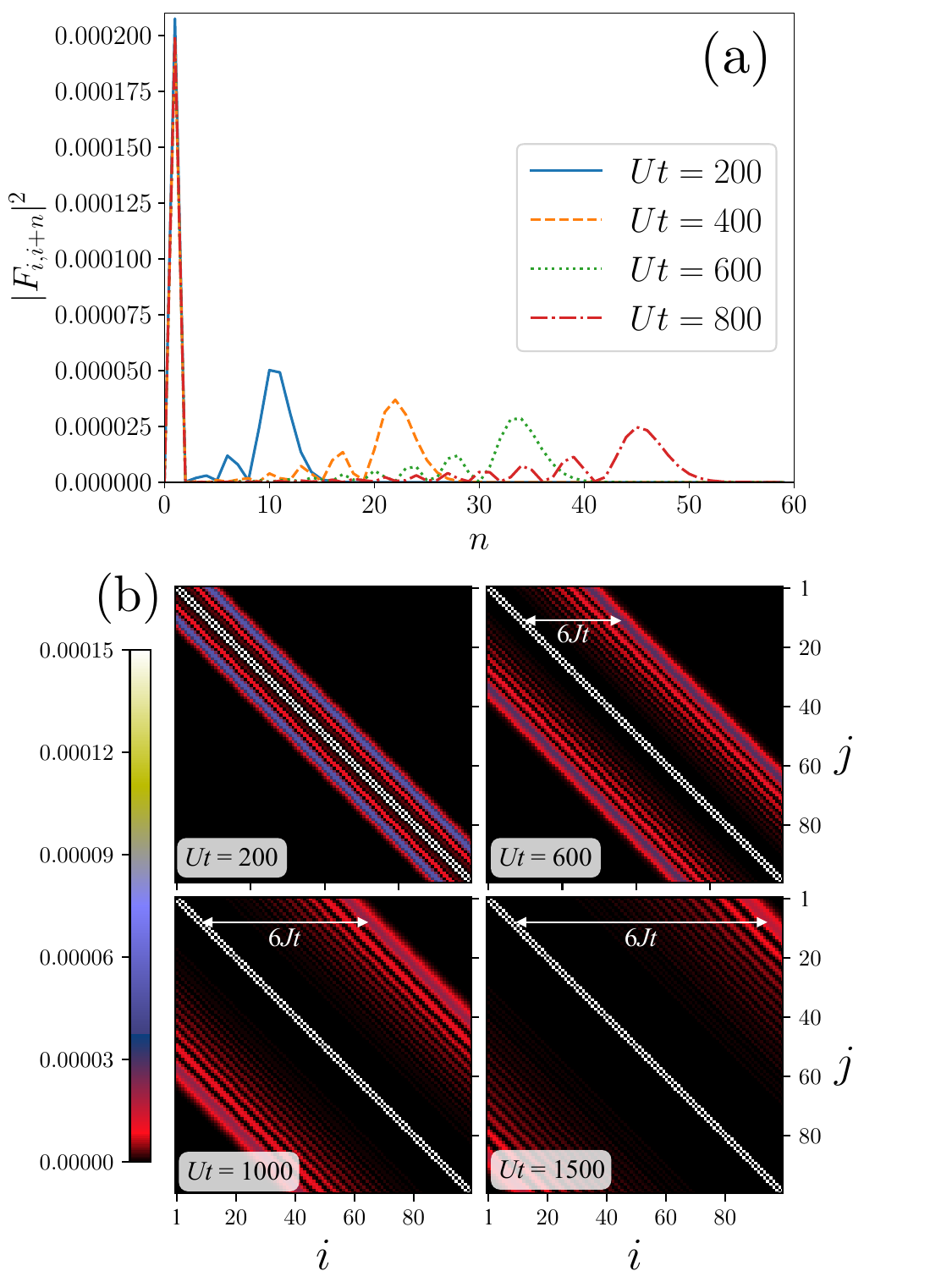}
\caption{(a) $|F_{i,i+n}(t)|^2$ as a function of $n$. (b) Colormap of $|F_{i,j}(t)|^2$. We set $J/U=0.01$ in both (a) and (b).
In (b), we take the subsystem size $L_{\mathrm{A}} = 100$.
}
\label{fig:F_ij_1500}
\end{figure}

As for the time evolution of the RE, it can be understood by studying the anomalous correlation function. Figure~\ref{fig:F_ij_1500} (a) shows $|\la \psi(t)|d_i h_{i+n}|\psi(t)\ra|^2=|F_{i,i+n}(t)|^2$ as a function of $n$. The propagating peaks in the figure correspond to those of the Bessel function ${\cal J}_n(6Jt)$ ($1\le n\le L_{\rm A}-1$) in Eq.~(\ref{eq:F_n}). Each of them describes a wave packet of entangled doublon-holon pairs induced by the quench.
The most dominant peak of ${\cal J}_n(6Jt)$ at $n\simeq6Jt$ represents pairs emitted at the initial time $t=0$. The sub-dominant propagating peaks represent pairs emitted at $t>0$. All the peaks propagate with the same velocity $v_{\rm pair}=6J$, which coincides with the maximum group velocity of a doublon-holon pair with opposite momenta given by
\begin{align}
v_{\rm pair}=\max_k[\partial_k(\epsilon_{d,k}+\epsilon_{h,-k})] \simeq 6J. 
\end{align}
This propagation speed of doublon-holon pairs has been experimentally confirmed by measuring correlation functions \cite{Cheneau-2012,Takasu-2020}.
These pairs decay in time as shown in Fig.~\ref{fig:F_ij_1500} (a) due to the factor $3Jt$ in the denominator of the second term of Eq.~(\ref{eq:F_n}).
\par
To understand the characteristic features of the long-time dynamics, namely, the linear growth for $6Jt<L_{\mathrm{A}}$ and the following saturation for $6Jt>L_{\mathrm{A}}$, we focus on the matrix $|F_{i,j}(t)|^2$, which is visualized in Fig.~\ref{fig:F_ij_1500} (b).
One finds that, when $6Jt<L_{\rm A}$, the contribution of pairs emitted at $t=0$ decreases as the size of pairs grows. It leads to the linear growth of $S_{\mathrm{A}}(t)$ for $6Jt<L_{\rm A}$. When $6Jt>L_{\rm A}$, pairs emitted at $t=0$ spread beyond the subsystem size and their contribution to $\norm{F}_{\rm F}^2$ vanishes. This results in the saturation of $S_{\rm A}$.
The smooth transition from the linear growth to the saturation of $S_{\rm A}$ is due to the contribution of the sub-dominant peaks, i.e., pairs emitted at $t>0$.
\par
In addition to the propagating peaks, there is a single localized peak at $n=1$ in Fig.~\ref{fig:F_ij_1500} (a), which corresponds to the subdiagonal elements in Fig.~\ref{fig:F_ij_1500} (b). It results in the second term in Eq.~(\ref{eq:RE_analytic}) and its height oscillates with the frequency $U$.
This localized peak represents doublon-holon pairs with unit separation spanning the boundary of subsystem A. They are excited by hopping of a boson to the nearest-neighbor sites. They repeat creation and annihilation with the frequency $U$ and eventually decay.
\par 
We compare the above quasiparticle picture and the one proposed by Calabrese and Cardy for quench dynamics of a general free-fermion model \cite{Calabrese-2005} to highlight our original results.
First of all, they proposed a quasiparticle picture for quench dynamics of a general free-fermion model, which is only valid in the space-time scaling limit ($t,L_\mathrm{A}\rightarrow\infty$ with $L_\mathrm{A}/t$ fixed). In contrast, our picture is derived microscopically and not restricted within the space-time scaling limit. In the present work, the quasiparticle picture is derived not only in the space-time scaling limit, but also in the short-time scale and/or small subsystems. We find, for example, that localized doublon-holon pairs with unit separation yield rapid oscillation of $S_\mathrm{A}$ in the short-time dynamics. In addition, our quasiparticle picture for the ground state is indeed not included in their picture. Furthermore, our picture has some remarkable features even in the space-time scaling limit that are absent in their picture. First, while the dynamics of entanglement entropy are described in terms of quasiparticle pairs emitted only at $t=0$ in their picture, we find that doublon-holon pairs emitted after the initial time also play crucial roles in the dynamics of the RE. In particular, the smooth transition from the linear growth to the saturation of $S_\mathrm{A}$ can be explained by their presence.
In addition, the second asymptotic form of the RE in Eq.~(\ref{eq:RE_asymptotic}) for $Jt\gg L_A$ can be explained by these pairs emitted at $t>0$.
Second, we find that the doublon-holon pairs decay as they propagate. This is also absent in their picture. 
Meanwhile, we confirm their predictions in the space-time scaling limit in our system. Their quasiparticle picture predicts that the entanglement entropy grows linearly up to $t\simeq L_\mathrm{A}/2$ (in units where the speed of elementary excitations is unity) and then is saturated to a value proportional to $L_\mathrm{A}$ \cite{Alba-2017}. 
This is indeed consistent with Fig.~\ref{fig:S_V_law} (b) and Eq.~\eqref{eq:RE_asymptotic}.
\par
In closing this section, we remark on the possibility of experimental verification of our results on the dynamics of the RE.
It may be difficult to experimentally confirm our predictions on the long-time dynamics of the RE in Fig.~\ref{fig:S_V_law} (b) due to the limitation of the lifetime of an atomic gas and the finite size effect. The short-time dynamics of the RE in Fig.~\ref{fig:S_V_law} (a) may be verified using the experimental setup in Refs.~\cite{Islam-2015,Kaufman-2016}.

\section{$n$th-order R\'enyi entropy}
\label{sec:nth-order_RE}
We have so far focused on the 2nd-order RE because it is experimentally accessible.
Meanwhile, from a theoretical point of view, it is desirable to extend our analysis to the $n$th-order RE. In particular, the 1st-order RE (von Neumann entanglement entropy) is a fundamental quantity that characterizes entanglement.
In this section, we calculate the $n$th-order RE for the ground state $|\mathrm{vac}\ra$ and the time-evolving state $|\psi(t)\ra$. 
\par
The $n$th-order RE for a reduced density matrix $\hat \rho_\mathrm{A}$ is defined as 
\begin{align}
    S^{(n)}_\mathrm{A}
    = 
    \frac{1}{1-n} \ln [\tr_{\rm A} (\hat{\rho}_\mathrm{A}^n)].
\end{align}
For a Gaussian state, using Eqs.~(\ref{eq:RDM_vac_main}) and (\ref{eq:ent_Ham_vac}), it can be written as
\begin{align}
    S_\mathrm{A}^{(n)}
    = 
    \frac{1}{1-n}
    \sum_{\alpha=1}^{2L_\mathrm{A}} \ln[f_\alpha^n+(1-f_\alpha)^n],
    \label{eq:Sn}
\end{align}
where $f_\alpha$ can be obtained from Eq.~\eqref{eq:M_diag_main}.
\par
Analogous to the 2nd-order RE, we expand $f_\alpha^n + (1-f_\alpha)^n$ in terms of $f_{\alpha} - f_{\alpha}^2 = O[(J/U)^2]$ for $n \geq 2$ as
\begin{align}
    f_\alpha^n + (1-f_\alpha)^n=1-n (f_\alpha-f_\alpha^2) +O[(J/U)^4].
    \label{eq:g_expansion}
\end{align}
We can easily verify the above relation by using the fact that $g^{(n)}_\alpha := f_\alpha^n + (1-f_\alpha)^n$ satisfies
\begin{align}
    g_\alpha^{(n+2)} = g^{(n+1)} - (f_\alpha-f_\alpha^2) g^{(n)}_\alpha.
\end{align}
As a result, we obtain the $n$th-order RE ($n\geq 2$)
\begin{align}
    S^{(n)}_\mathrm{A} 
    &=
    \frac{n}{n-1} \sum_{\alpha=1}^{2L_\mathrm{A}} (f_\alpha-f_\alpha^2) 
    +O[(J/U)^4].
\end{align}
Using Eq.~\eqref{eq:f_vac}, it is proportional to the 2nd-order RE as
\begin{align}
    S_\mathrm{A}^{(n)}
    &= 
    \frac{n}{2(n-1)}S_\mathrm{A}^{(2)} +O[(J/U)^4].
   \label{eq:Sn_S2_relation}
\end{align}
\par
We need to calculate $S^{(1)}_\mathrm{A}$ numerically, because Eq.~(\ref{eq:g_expansion}) does not hold for $n=1$.
Taking the limit $n\rightarrow 1$ in Eq.~(\ref{eq:Sn}), we obtain 
\begin{align}
    S^{(1)}_\mathrm{A}
    = 
    -\sum_{\alpha=1}^{2L_\mathrm{A}}
    [(1-f_\alpha) \ln(1-f_\alpha)+f_\alpha\ln(f_\alpha)]. 
    \label{eq:EE}
\end{align}
We evaluate $f_\alpha$ by numerically diagonalizing the matrix $M$ and plugging them into the above equation. 
\par
\begin{figure}[!t]
    \includegraphics[width=0.5\textwidth]{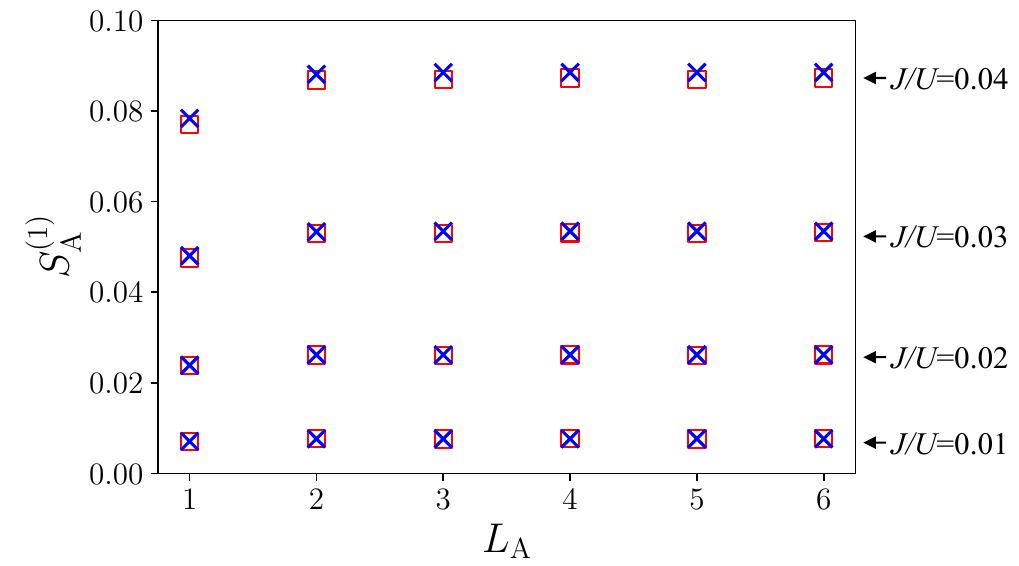}
    \caption{
    1st-order RE for the ground state $|\mathrm{vac}\ra$ as a function of $L_{\rm A}$.
    The crosses represent the results obtained from Eq.~(\ref{eq:EE}), where $f_\alpha$ are obtained by numerically diagonalizing the matrix $M$.
    The squares represent the numerical results of the iTEBD calculation.
    }
    \label{fig:EE_vac}
\end{figure}
\begin{figure}[!t]
    \includegraphics[width=0.4\textwidth]{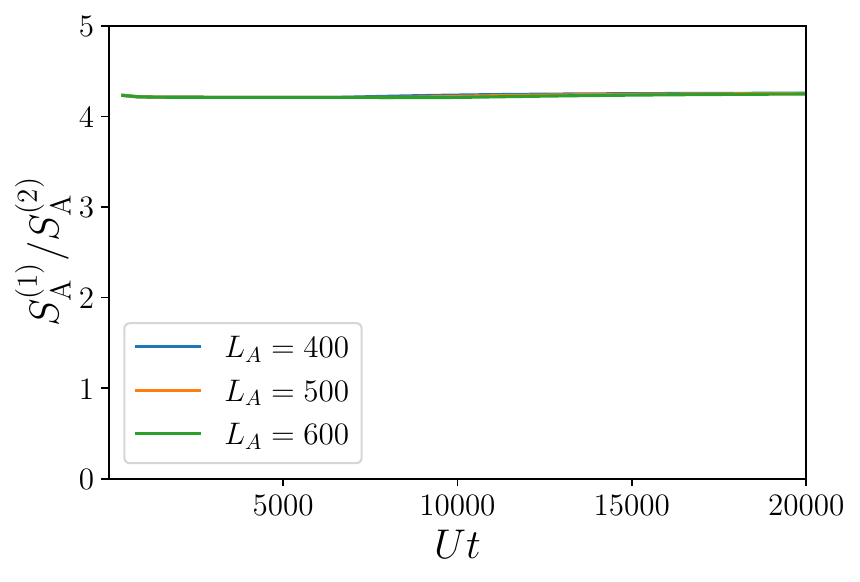}
    \caption{Time evolution of the ratio of the 1st-order RE to the 2nd-order one. We set $J/U=0.01$. We note that all lines take almost the same value.} 
    \label{fig:EE_t}
\end{figure}

Figure~\ref{fig:EE_vac} shows the 1st-order RE for the ground state $|\mathrm{vac}\ra$ as a function of $L_\mathrm{A}$.
$S^{(1)}_\mathrm{A}$ is almost independent of $L_\mathrm{A}$, which implies that the 1st-order RE follows the area-law scaling.
Figure~\ref{fig:EE_t} shows the ratio of the 1st-order RE to the 2nd-order RE as a function of time. The ratio is almost constant in time and shows very little dependence on subsystem size $L_A$.
The RE in arbitrary order $n\ge 1$ thus exhibits qualitatively the same behavior as the 2nd-order RE in the dynamics as well as in the ground state. Its behavior indeed reflects the quasiparticle picture described in Sec.~\ref{sec:QP}.

\section{Validity of the effective theory}
\label{sec:validity_of_the_effective_theory}

We have studied the RE within the effective theory. The effective theory (\ref{eq:H_eff}) is integrable whereas the original BHM (\ref{eq:BHM}) and the effective Hamiltonian with the projection operator (\ref{eq:H_PHeffP}) are non-integrable. This difference may induce deviations in REs in the region away from the deep MI regime. Therefore, it is important to clarify how large $J/U$ can be for the effective theory to be qualitatively valid. To this end, we compare the REs calculated by the effective theory and the iTEBD algorithm.
\par
\begin{figure}
    \includegraphics[width=0.49\textwidth]{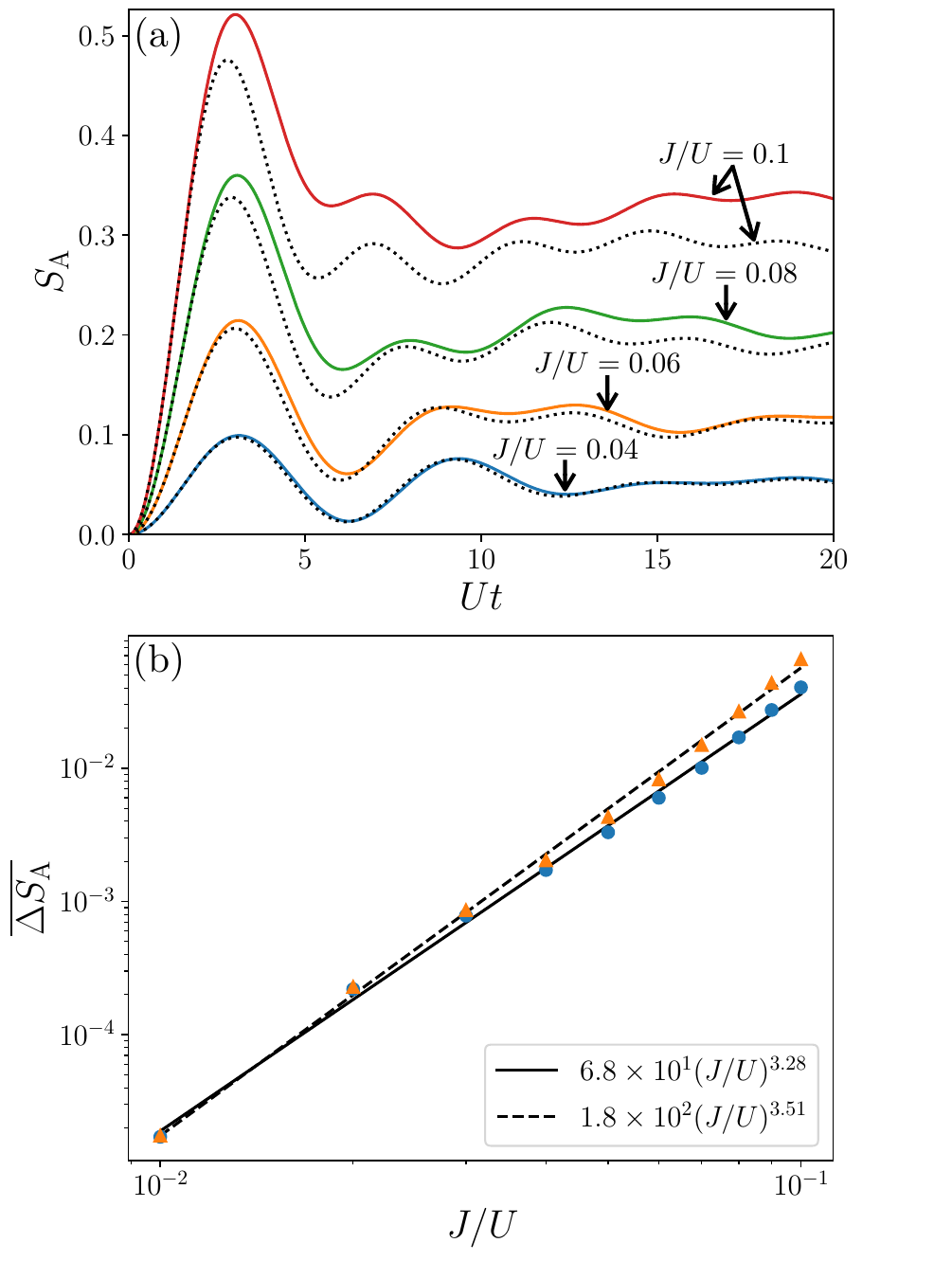}
    \caption{(a) Time evolution of the RE for several values of $J/U$. The solid and dotted lines represent $S_\mathrm{A}^{\rm eff}$ and $S_\mathrm{A}^\mathrm{iTEBD}$, respectively. We set $L_\mathrm{A}=2$. (b) Log-log plot of $\overline{\Delta S_\mathrm{A}}$ versus $J/U$. The dots and triangles represent the results for $L_\mathrm{A}=1$ and 2, respectively.
    The solid and dashed lines show power-function fits of $\overline{\Delta S_\mathrm{A}}$ for $L_\mathrm{A}=1$ and 2, respectively.}
    \label{fig:dS}
\end{figure}

 
Figure~\ref{fig:dS} (a) shows a comparison of their time evolution.
It clearly shows that the effective theory overestimates the RE. Furthermore, the overestimation increases in time during the linear growth and stops increasing after it.
This overestimation may arise from the approximation in deriving the effective Hamiltonian (\ref{eq:H_eff}), in which we ignore the projection operator and allow unphysical double occupancy of a doublon and a holon.
This approximation may result in overestimation of the number of doublon-holon pairs and accordingly the RE, because doublon-holon pairs are responsible for the RE.
The overestimation of the number of doublon-holon pairs is expected to increase during the linear growth of the RE, because doublon-holon pairs keep generating in this period as we observe in Sec.~\ref{sec:QP}. 
This may result in the increase of the overestimation in the linear growth part in Fig.~\ref{fig:dS} (a).
Note that the projection operator introduces non-integrable effects that correspond to interactions of bogolons.
\par
Figure~\ref{fig:dS} (a) also shows that the overestimation increases as $J/U$ increases.
To understand it quantitatively, we calculate time-averaged overestimation
\begin{align}
    \overline{\Delta S_\mathrm{A} }
    =
    \frac{1}{M}
    \sum_{n=1}^{M}
    |S_\mathrm{A}^{\rm eff}(n \Delta t)-S_\mathrm{A}^\mathrm{iTEBD}(n \Delta t)|,
\end{align}
where $S_\mathrm{A}^{\rm eff}$ and $S_\mathrm{A}^\mathrm{iTEBD}$ denote the RE obtained by the effective theory and iTEBD calculations, respectively, and $M=t_\mathrm{max}/\Delta t$ the number of data points in the interval $0\le t \le t_\mathrm{max}$. We set $U t_{\rm max}= 20$.
Figure~\ref{fig:dS} (b) shows a log-log plot of $\overline{\Delta S_\mathrm{A}}$ versus $J/U$. Since the time-averaged overestimation scales as $\overline{\Delta S_\mathrm{A}} \propto (J/U)^{3.2\text{--}3.6}$, the effective theory is valid as long as $(J/U)^3 \ll 1$. Since our quasiparticle picture is valid within the effective theory, it is also valid under the same condition.
This condition is consistent with a previous study based on the analysis of a density correlation function \cite{Barmettler-2012}, in which it is claimed that the effective theory is valid as long as $J/U\leq 1/8$.
Note that we confirmed that the power of $J/U$ is larger than 3 regardless of the values of $\Delta t$ and/or $t_\mathrm{max}$.

\section{summary}\label{sec:summary}
In summary, we have investigated the time evolution of the RE for bosons in a one-dimensional optical lattice when the system undergoes a quench from the MI limit to the strongly correlated MI regime. 
Developing the effective theory, we have derived a direct relation between the RE and correlation functions associated with doublon and holon excitations. Using this relation, we have calculated the RE analytically and obtained a physical picture, both in the ground state and during time evolution, in terms of entangled doublon-holon pairs. 
Our quasiparticle picture for the dynamics of the RE introduces some features that are absent in previous studies on the dynamics of entanglement entropy in free-fermion models.
This quasiparticle picture provides valuable insight into the quench dynamics of strongly correlated quantum many-body systems.

\acknowledgements
We thank I. Danshita, R. Kaneko, K. Sugiyama, and Y. Takeuchi for fruitful discussions. ST thanks D. Huerga and R. Raussendorf for stimulating discussions. SY is supported by Grant-in-Aid for JSPS Fellows (Grant No.\,JP22J22306). DK was supported by JST CREST (Grant No.~JPMJCR1673) and JST FOREST (Grant No.~JPMJFR202T). RY is partially supported by a Grant-in-Aid of MEXT for Scientific Research KAKENHI (Grant Nos.~19K14616 and 20H01838). ST is supported by the Japan Society for the Promotion of Science Grant-in-Aid for Scientific Research (KAKENHI Grant No.~19K03691).

\appendix

\section{Derivation of the effective Hamiltonian}
\label{App:Heff}

We derive the effective Hamiltonian Eq.~(\ref{eq:H_eff}) from the BHM in this Appendix.
When $J/U$ is sufficiently small, the system can be mapped to a spin-1 system as it can be effectively described within the truncated Hilbert space spanned by $|0\ra_j$, $|1\ra_j$, and $|2\ra_j$, where they correspond to the spin-1 states: $\ket{2}_j=\ket{S_z=1}_j$, $\ket{1}_j=\ket{S_z=0}_j$, and $\ket{0}_j=\ket{S_z=-1}_j$ \cite{Altman-2002}. Introducing the spin-1 operators 
\begin{gather}
\h S_j^+ = \sqrt 2 (\ket{S_z=1}_j\bra{S_z=0}_j+\ket{S_z=0}_j\bra{S_z=-1}_j),\\
\h S_j^- = \sqrt 2 (|S_z=0\ra_j\la S_z=1|_j+|S_z=-1\ra_j\la S_z=0|_j),\\
\h S_j^z = |S_z=1\ra_j\la S_z=1|_j - |S_z=-1\ra_j\la S_z=-1|_j, \label{eq:S_z}
\end{gather}
the BHM (\ref{eq:BHM}) can be written as
\begin{align}
\h H 
=&
-\frac{J}{2} \sum_j [ \h S_j^+ \h S_j^z \h S_{j+1}^z \h S_{j+1}^- + 2 \h S_j^z \h S_j^+ \h S_{j+1}^- \h S_{j+1}^z 
\nn
& -\sqrt 2 (\h S_j^+ \h S_j^z \h S_{j+1}^- \h S_{j+1}^z + \h S_j^z \h S_j^+ \h S_{j+1}^z \h S_{j+1}^-) + {\rm H.c.} ]
\nn
& + \frac{U}{2} \sum_j \h S_j^z {}^2
+ \frac{U}{2}\sum_j \h S_j^z
. \label{eq:BHM_spin}
\end{align}
We note that the same mapping was used in Ref.~\cite{Huber-2007} and our Hamiltonian in Eq.~(\ref{eq:BHM_spin}) can be obtained by setting the average filling $n_0=1$ in Eq.~(A5) in Ref.\,\cite{Huber-2007}.
Since $\sum_j \h S_j^z = \sum_j (\h n_j-1) = N-L$ is a constant, we neglect the last term in Eq.~(\ref{eq:BHM_spin}). 
\par
We introduce the pseudo-fermion operators $\bar d_j$ and $\bar h_j$ by the generalized Jordan-Wigner transformation \cite{Batista-2001}
\begin{gather}
\h S_j^+ = \sqrt 2 (\bar d_j^\dg + \bar {h}_j) \h K_j, \label{eq:JW_1}\\
\h S_j^- = \sqrt 2 \hat K_j (\bar d_j + \bar h_j^\dg),\label{eq:S_j^-} \\
\h S_j^z = \bar n_{jd} -  \bar n_{jh}, \label{eq:S_j^z}
\end{gather}
where $\bar n_{jd} = \bar d_j^\dg \bar d_j$, $\bar n_{jh} = \bar h_j^\dg \bar h_j$, and $\h K_j = \prod_{i<j}(1-2\bar n_{id})(1-2\bar n_{ih})=\prod_{i<j}(-1)^{(\h S_i^z)^2}$ is the string operator. The inverse transformation is given by 
\begin{gather}
\bar d_j = \frac{1}{\sqrt 2}\h K_j \h S_j^- \h S_j^z, \label{eq:JW_inv_1}\\
\bar h_j = -\frac{1}{\sqrt 2}\h K_j \h S_j^+ \h S_j^z. \label{eq:JW_inv_2} 
\end{gather}
$\bar d_j$ and $\bar h_j$ are annihilation operators of ``particle" and ``hole" excitations, respectively, because $\bar d_j|1\ra_j = \bar h_j|1\ra_j=0$, $|2\ra_j=\h K_j\bar d_j^\dg|1\ra_j\propto \bar d_j^\dg|1\ra_j$, and $|0\ra_j=\h K_j\bar h_j^\dg|1\ra_j \propto \bar h_j^\dg|1\ra_j$. However, since their multiple occupation on the same site is prohibited ($\bar d_j^\dg \bar h_j^\dg = 0$), they do not obey the fermionic anti-commutation relations. Note that the string operator $\h K_j$ becomes a phase factor that depends on $(\h S_i^z)^2$ ($1\leq i\leq j-1$). 
\par
We further introduce the annihilation operators $\h d_j$ and $\h h_j$ as
\begin{gather}
\bar { d}_j = (1-\h n_{jh}) \h d_j, \label{eq:d}\\
\bar { h}_j = (1-\h n_{jd}) \h h_j, \label{eq:h}
\end{gather}
where $\h n_{jd}=\h d_j^\dg \h d_j$ and $\h n_{jh}=\h h_j^\dg \h h_j$.
They represent fermionic excitations as they satisfy the usual anti-commutation relations: $\{\h d_i,\h d_j^\dg\}=\{\h h_i,\h h_j^\dg\}=\delta_{i,j}$, $\{\h d_i,\h d_j\} = \{\h h_i,\h h_j\}=\{\h d_i,\h h_j\}=\{\h d_i,\h h_j^\dg\}=0$.
We obtain $\ket{0}_j=\ket{n_d=0,n_h=1}_j$, $|1\ra_j=|n_d=0,n_h=0\ra_j$, and $\ket{2}_j=\ket{n_d=1,n_h=0}_j$ from $\h S_j^z = \h n_{jd} -\h n_{jh}$. $\h d_j$ and $\h h_j$ thus describe fermionic particle and hole excitations, respectively. We refer to them as ``doublon" and ``holon" in this paper. Note that $|1\ra_j=|n_d=0,n_h=0\ra_j$ is the vacuum of $\h d_j$ and $\h h_j$. $|n_d=1,n_h=1\ra_j$ has no corresponding Fock states of original bosons.
\par
Substituting Eqs.~(\ref{eq:JW_1})-(\ref{eq:S_j^z}) into Eq.~(\ref{eq:BHM_spin}), we obtain an intermediate form of the Hamiltonian
\begin{align}
\hat{H}
=& 
-J \sum_j [ 2 \bar{d}_j^\dag \bar{d}_{j+1} + \bar{h}_{j+1}^\dag \bar{h}_j \nn
&+ \sqrt 2 (\bar{d}_j^\dag \bar{h}_{j+1}^\dag - \bar{h}_j \bar{d}_{j+1})+{\rm H.c.}] \nn
&+ \frac{U}{2} \sum_j (\bar{d}_j^\dag \bar{d}_j + \bar{h}_j^\dag \bar{h}_j).
\end{align}
In order to take into account the condition $\bar d_j^\dg \bar h_j^\dg = 0$, it is convenient to introduce the projection operator $\hat{P} = \prod_j (1-\hat{n}_{jd}\hat{n}_{jh})$ which eliminates fictitious states $|n_d=1,n_h=1\ra_j$.
Using $\bar{d}_j^\dag \bar{d}_{j+1} = \hat{P} \hat{d}_j^\dag \hat{d}_{j+1} \hat{P}$ and so on, we obtain the Hamiltonian in Eqs.\,(\ref{eq:H_PHeffP}) and (\ref{eq:H_eff}).

\section{Single-site reduced density matrix for the time-evolving state}
\label{app:rho_A_single}

In this Appendix, we derive the reduced density matrix for a single site Eq.~\eqref{eq:rdm_effective_explicit}.
First, we express the operators $|\tau\rangle_j \langle \tau'|_j~(\tau,\tau' = \{(0,0),(1,0),(0,1),(1,1)\})$ in terms of the annihilation and creation operators of doublons and holons.
Diagonal elements of $|\tau\rangle_j \langle \tau'|_j$ are given by
\begin{align}
|0,0 \rangle_j \langle 0,0|_j =& (1 - \hat{d}^{\dag}_{j} \hat{d}_{j}) (1 - \hat{h}^{\dag}_{j} \hat{h}_{j}),
\\
|1,0 \rangle_j \langle 1,0|_j =& \hat{d}^{\dag}_{j} \hat{d}_{j} (1 - \hat{h}^{\dag}_{j} \hat{h}_{j}),
\\
|0,1 \rangle_j \langle 0,1|_j =& (1 - \hat{d}^{\dag}_{j} \hat{d}_{j}) \hat{h}^{\dag}_{j} \hat{h}_{j},
\\
|1,1 \rangle_j \langle 1,1|_j =& \hat{d}^{\dag}_{j} \hat{d}_{j} \hat{h}^{\dag}_{j} \hat{h}_{j}.
\end{align}
Off-diagonal elements are given by
\begin{align}
|1,0 \rangle_j \langle 0,0|_j =& \hat{d}^{\dag}_{j} (1 - \hat{h}_{j} \hat{h}_{j}),\label{eq:B5}
\\
|0,1 \rangle_j \langle 0,0|_j =& (1 - \hat{d}^{\dag}_{j} \hat{d}_{j}) \hat{h}^{\dag}_{j},
\\
|1,1 \rangle_j \langle 0,0|_j =& \hat{d}^{\dag}_{j} \hat{h}^{\dag}_{j},
\\
|0,1 \rangle_j \langle 1,0|_j =& \hat{d}_{j} \hat{h}^{\dag}_{j},
\\
|1,1 \rangle_j \langle 1,0|_j =& \hat{d}^{\dag}_{j} \hat{d}_{j} \hat{h}^{\dag}_{j},
\\
|1,1 \rangle_j \langle 0,1|_j =& \hat{d}^{\dag}_{j} \hat{h}^{\dag}_{j} \hat{h}_{j}.
\label{eq:B10}
\end{align}
Other off-diagonal elements can be obtained by taking Hermitian conjugation of Eqs.~(\ref{eq:B5})-(\ref{eq:B10}).

We calculate each element of the reduced density matrix using Eq.~(\ref{eq:rdm_tomography}).
Explicit calculations show that the expectation values of the off-diagonal elements of $|\tau\rangle_j \langle \tau'|_j$ vanish.
For diagonal elements, due to the fact that the time-evolving state $|\psi(t)\rangle$ is a Gaussian state, as discussed in Sec.~\ref{sec:RE_L_A>1}, we can evaluate correlation functions using the Wick decomposition.
Noting that $\langle \psi(t) | \hat{d}^{\dag}_{j} \hat{h}^{\dag}_{j} | \psi(t) \rangle = \langle \psi(t) | \hat{d}_{j} \hat{h}_{j} | \psi(t) \rangle = 0$, we can express the reduced density matrix in the form of Eq.~\eqref{eq:rdm_effective_explicit}.

\section{Derivation of Eq.\,(\ref{eq:S_Bessel_omg})}
\label{app:Fourier}
In this Appendix, we derive Eq.~(\ref{eq:S_Bessel_omg}). Substituting Eq.~(\ref{eq:S_Bessel}) into Eq.~(\ref{eq:S_omg_1}), we obtain 
\begin{align}
    \bar S_\mathrm{A}(\omega)
    &\simeq 
    -32(J/U)^2
    \int_{-\infty}^\infty\dd t
    e^{\im \omega t}
    \frac{\mathcal{J}_1(6Jt)}{3Jt}\cos(Ut).
    \label{eq:S_fourier}
\end{align}
Here, we have neglected terms of order $O[(J/U)^4]$.
We have also ignored the time-independent terms by assuming $\omega>0$. Using the identity 
\begin{align}
    \frac{\mathcal{J}_1(6Jt)}{3Jt}
    &= 
    \frac{1}{\pi}
    \int_{-\pi}^\pi \dd k e^{-6\im Jt\cos(k)} \sin^2(k),
\end{align}
Eq.\,(\ref{eq:S_fourier}) can be written as 
\begin{align}
    \bar S_A(\omega)
    \simeq&
    -\frac{16J^2}{\pi U^2}
    \int_{-\pi}^\pi \dd k 
    \sin^2(k) 
    \nonumber \\
    &\times
    \sum_{\sigma = \pm}
    \int_{-\infty}^\infty \dd t
    e^{\im [-6J\cos(k)+\omega + \sigma U]t}
    \nn
    =&
    -\frac{32J^2}{U^2}
    \int_{-\pi}^\pi \dd k 
    \sin^2(k) 
    \sum_{\sigma = \pm}
    \delta(\omega + \sigma U-6J\cos(k))
    \nn
    = &
    -\frac{32J}{3U^2} \sum_{\sigma = \pm}
    \sqrt{1-\qty(\frac{\omega+\sigma U}{6J})^2}\theta(6J-|\omega+\sigma U|).
\end{align}
From the condition $J \ll U$, we ignore the $\sigma = +$ contribution. Then, we obtain Eq.\,(\ref{eq:S_Bessel_omg}).

\section{Derivation of Eq.~(\ref{eq:rho_A_quad})}\label{App:rho_A}

We derive Eq.~(\ref{eq:rho_A_quad}) in this Appendix. The following arguments are based on Ref.~\cite{Frerot-2015}. 
\par 
Let us consider a many-body correlation function $\la \psi(t) |\h a_{i_1}\h a_{i_2},...,\h a_{i_{2n}}|\psi(t)\ra$, where $\h a_{i} \in\{\h d_i,\h h_i,\h d_i^\dg,\h h_i^\dg\}$. In the Heisenberg picture, it can be written as 
\begin{align}
\la \psi_0| \check a_{i_1}(t) \check a_{i_2}(t)...\check a_{i_{2n}}(t)|\psi_0\ra, \label{eq:many_corr}
\end{align}
where $\check a_i(t) = e^{\im H_{\rm eff}t} \h a_i e^{-\im H_{\rm eff}t}$. 
Since $\h H_{\rm eff}$ is a quadratic Hamiltonian of $\{\h a_i\}$, $\check a_i(t)$ can be expressed in a linear combination of $\{\h a_i\}$. Given that $|\psi_0\ra$ is the vacuum state of $\h d_i$ and $\h h_i$, Eq.~(\ref{eq:many_corr}) can be decomposed into one-body correlation functions by Wick's theorem \cite{Wick-1950}. We thus obtain 
\begin{gather}
\la \psi_0|
\check a_{i_1}(t)\check a_{i_2}(t),...,
\check a_{i_{2n-1}}(t) \check a_{i_{2n}}(t)
|\psi_0 \ra = 
\nn
\sum_{\sigma \in P} {\rm sgn}(\sigma) \la \psi_0 |\check a_{j_1}(t)\check a_{j_2}(t)|\psi_0\ra 
\la \psi_0|\check a_{j_3}(t)\check a_{j_4}(t)|\psi_0\ra
\dots 
\nn
\times 
\la \psi_0 |
\check a_{j_{2n-3}}(t) \check a_{j_{2n-2}}(t)
|\psi_0\ra
\la \psi_0 |
\check a_{j_{2n-1}}(t) \check a_{j_{2n}}(t)
|\psi_0\ra, \label{eq:wick_1}
\end{gather} 
where $P$ is the set of permutations $(i_1,i_2,...,i_{2n})\rightarrow(j_1,j_2,...,j_{2n})$ satisfying $j_{k-1}<j_k$ and $j_1<j_3<...<j_{2n-1}$. Returning to the Scrh$\ddot{\rm o }$dinger picture, Eq.~(\ref{eq:wick_1}) becomes 
\begin{gather}
\la \psi(t)|
\h a_{i_1}\h a_{i_2},...,
\h a_{i_{2n-1}} \h a_{i_{2n}}
|\psi(t) \ra = \nn
\sum_{\sigma \in P} {\rm sgn}(\sigma) \la \psi(t) |
\h a_{j_1} \h a_{j_2} |\psi(t)\ra 
\la \psi(t)|\h a_{j_3} \h a_{j_4}|\psi(t)\ra
\dots
\nn
\times 
\la \psi(t) |
\h a_{j_{2n-3}} \h a_{j_{2n-2}}
|\psi(t)\ra
\la \psi(t) |
\h a_{j_{2n-1}} \h a_{j_{2n}}
|\psi(t)\ra. \label{eq:wick}
\end{gather}
If the many-body correlation function (\ref{eq:many_corr}) concerns the degrees of freedom in subsystem A, i.e., $\{i_1,i_2,...,i_{2n}\}\in {\rm A}$, all the correlation functions in Eq.~(\ref{eq:wick}) can be calculated by the reduced density matrix for A. In this case, Eq.~(\ref{eq:wick}) can be written as 
\begin{gather}
\tr_{\rm A}(\h \rho_{\rm A} \h a_{i_1}\h a_{i_2},...,\h a_{i_{2n}})
= \sum_{\sigma \in P} {\rm sgn}(\sigma)
\tr_{\rm A}(\h \rho_{\rm A} \h a_{j_1}\h a_{j_2}) \nn
\times
\tr_{\rm A}(\h \rho_{\rm A} \h a_{j_3}\h a_{j_4})
...\tr_{\rm A}(\h \rho_{\rm A} \h a_{j_{2n-1}}\h a_{j_{2n}}). \label{eq:wick_2}
\end{gather}
Equation (\ref{eq:wick_2}) shows that the Bloch-De Dominicis theorem \cite{Matsubara-1955} can be applied to the correlation functions evaluated by $\h \rho_{\rm A}$. It follows that $\h \rho_{\rm A}$ is a thermal state of a quadratic Hamiltonian of $\h d_j$ and $\h h_j$ ($j\in {\rm A}$). We thus obtain Eq.~(\ref{eq:rho_A_quad}).

\bibliography{ref} 

\end{document}